\documentclass{article}

\usepackage{arxiv}
\usepackage{amsmath}
\usepackage[utf8]{inputenc} 
\usepackage[T1]{fontenc}    
\usepackage{hyperref}       
\usepackage{url}            
\usepackage{booktabs}       
\usepackage{amsfonts}       
\usepackage{nicefrac}       
\usepackage{microtype}      
\usepackage{lipsum}
\usepackage{graphicx}
\usepackage{subcaption}
\usepackage{caption}
\usepackage[utf8]{inputenc} 
\usepackage{setspace}

\title{Who Gets the Job and How are They Paid? \\ Machine Learning Application on H-1B Case Data}

\author{
  Barry Shikun Ke\thanks{This paper is built upon the research project for APMA4903: Seminar in Applied Math at Columbia University. We thank Professor Chris Wiggins for helpful comments and guidance for the project. For replication code, data, and presentation slides please visit \texttt{https://github.com/BarryKeee/APMA4903}} \\
  Applied Mathematics\\
  Columbia University\\
  \\
   \And
 Angela Qiao\\
  Applied Mathematics\\
  Columbia University
}

\begin{document}
\maketitle

\begin{abstract}
In this paper, we use machine learning techniques to explore the H-1B application dataset disclosed by the Department of Labor (DOL), from 2008 to 2018, in order to provide more stylized facts of the international workers in US labor market. We train a LASSO Regression model to analyze the impact of different features on the applicant's wage, and a Logistic Regression with L1-Penalty as a classifier to study the feature's impact on the likelihood of the case being certified. Our analysis shows that working in the healthcare industry, working in California, higher job level contribute to higher salaries. In the meantime, lower job level, working in the education services industry and nationality of Philippines are negatively correlated with the salaries. In terms of application status, a Ph.D. degree, working in retail or finance, majoring in computer science will give the applicants a better chance of being certified. Applicants with no or an associate degree, working in the education services industry, or majoring in education are more likely to be rejected.
\end{abstract}


\onehalfspacing

\section{Introduction}
The United States has always been attractive to international students due to its welcoming culture, quality education and a strong job market. In 2017, there were 1.21 million international students in the country, around 25\% of international students worldwide. After graduation, some of them will choose to stay in the country and work for the U.S. firm.

For these foreign-born professionals, the first week of April is an extremely stressful time as the companies they work for are rushing to file their H-1B visa applications. Later in April, a random lottery will choose less than a half of the applicants and they are allowed to temporarily work in the country. Employers must attest, on a labor  condition  application  (LCA)  certified  by  the  Department  of  Labor  (DOL),  that  employment  of  the  H-1B  worker  will  not  adversely  affect  the  wages  and  working  conditions  of  similarly  employed  U.S.  workers.

Research have been done on analyzing the overall impact of H-1B policy. However, we've noticed that Labor  Condition  Application  ("LCA")  disclosure data \footnote{https://www.foreignlaborcert.doleta.gov/performancedata.cfm\#\#dis} from the  U.S. Department of Labor provides includes comprehensive information regarding wage, industry, application decision, etc. 

In this paper, we will first do data exploration and look at how wages and number of applications differ by factors such as job sectors, states, and citizenship. We will then use Lasso Regression to look at how different factors impact the wage. We will also conduct a logistic regression to predict the status of the application (certify/deny) based on the profile (wage, sector, etc). 

Our analysis shows that working in the healthcare industry, working in California, higher job level contribute to higher salaries. In the meantime, lower job level, working in the education services industry and nationality of Philippines are negatively correlated with the salaries. In terms of application result, a Ph.D. degree, working in retail or finance, majoring in computer science will give the applicants a better chance of being certified. Applicants with no or an associate degree, working in the education services industry, or majoring in education are more likely to be rejected.

To the best of our knowledge, the H-1B dataset has not been deeply explored by statisticians or economists. One recent literature in economics (\cite{econ}) founds a negative effect of H-1B cap restriction on H-1B hiring by for-profit firms but does not change the hiring of US-born workers, which implies a low degree of substitution in the labor market between foreign and domestic workers. They also find redistribution of H-1Bs towards computer-related occupations, Indian-born workers and firms with intensive H-1B usage history, which are confirmed by our analysis. However, many cross-sectional features in the dataset are not explored, and we would like to provide more stylized facts of foreign labor supply and demand in the US by exploring the H-1B applications. Using the results, we are able to know more about which factors impact wage distribution and the application decision. Domestic and foreign workers are able to gain reference to the average wage of different levels, and further research could be done to investigate whether there exists wage discrimination against foreign workers. Students and firms can use the status prediction model to estimate the probability of being certified by the Department of Labor. 

\section{Exploratory Data Analysis}\label{EDA}
We will be looking at wage and application data from 2008 to 2015. The variables included are wage, date of application, employer name, location, economic sector, job title, and citizenship. After 2015, we have additional variables including total number of employees, the founding year of the firm, education level, university, major, and prior working experiences.

\begin{figure}
  \centering
    \begin{subfigure}[b]{0.45\textwidth}
        \includegraphics[width=\linewidth]{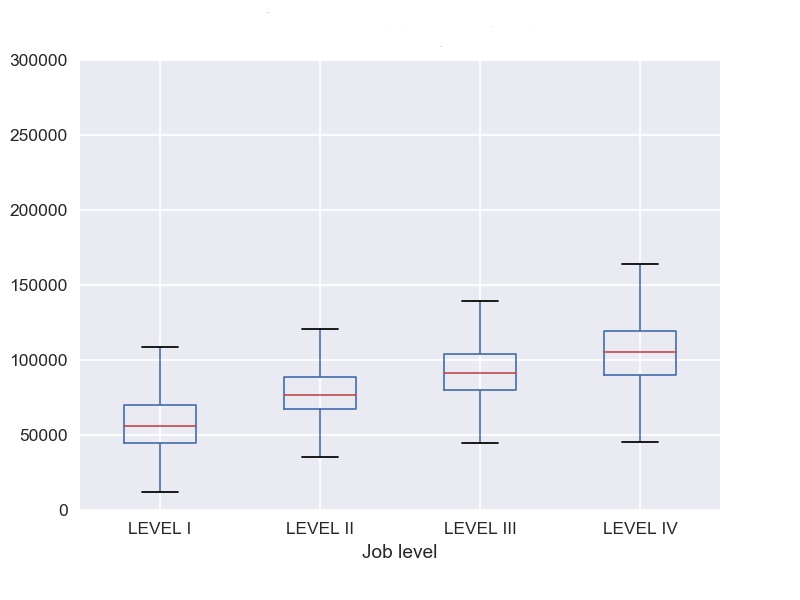}
        \caption{Box plot of wages by job level}
    \end{subfigure}
    \begin{subfigure}[b]{0.45\textwidth}
        \includegraphics[width=\linewidth]{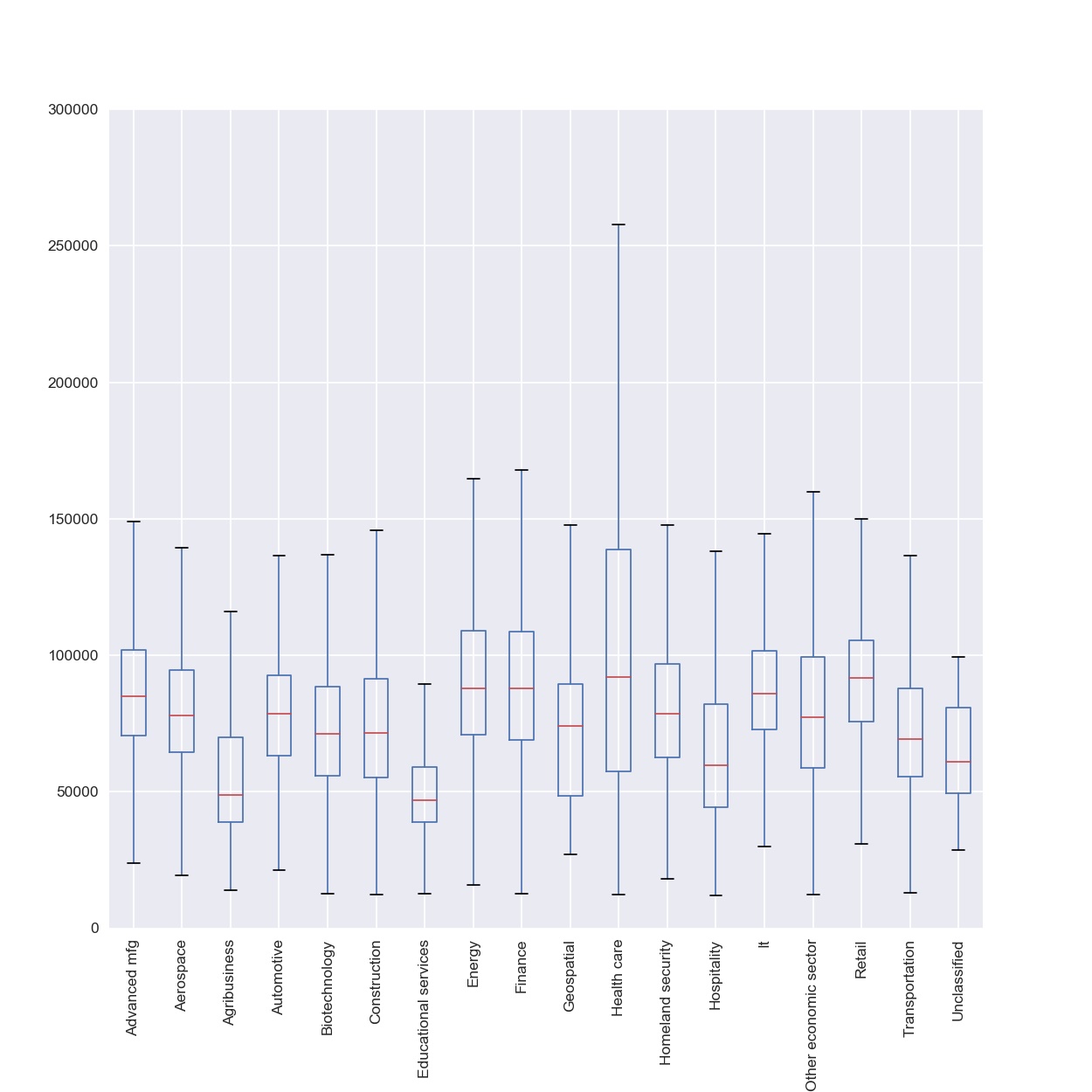}
        \caption{Box plot of wages by sector}
    \end{subfigure}
    \\
    ~ 
    \begin{subfigure}[b]{0.5\textwidth}
        \includegraphics[width=\linewidth]{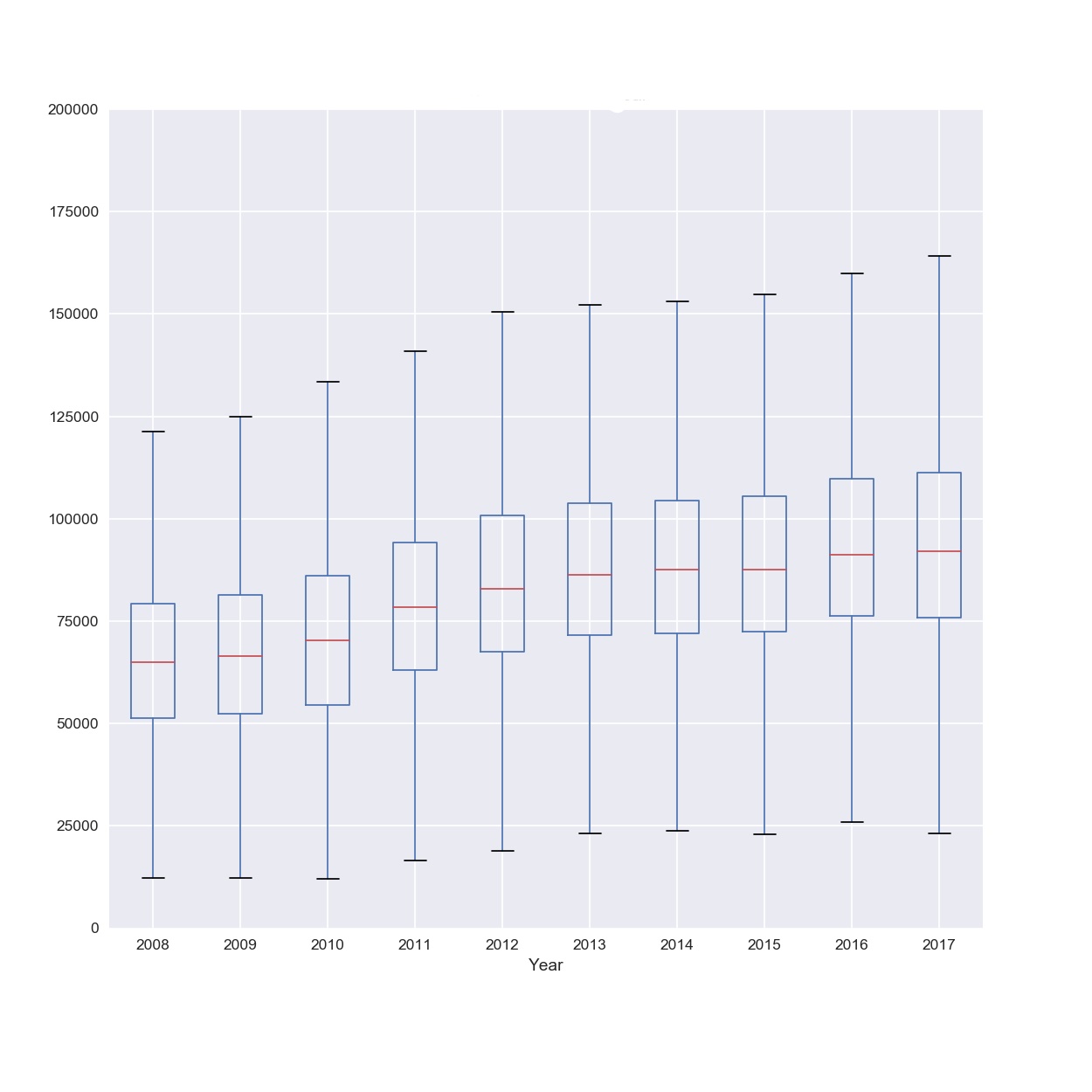}
        \caption{Box plot of wages by year}
    \end{subfigure}
    \caption{Box plot of wages}
\end{figure}

We first do a box plot of wages by job level. There are four levels in total and the medium wage increases by job level. Level IV has the highest range and level II has the lowest range. In terms of sectors, people who work in healthcare, retail, finance, and information technology earn the highest wages, while as people who work in agriculture and education service earn the lowest. People who work in healthcare sector also have the highest difference in wages. The wages increase by year and the biggest jump is from 2010 to 2011 as the country has just recovered from the financial crisis in 2008. It has grown steadily after 2011 and has been stable in recent years.

Indians file the greatest number of applications and account for around 75 percent of the total applicants, followed by Chineses, Canadians and Koreans. An US-based Indian technology company Cognizant Technology Solutions, along with Microsoft, Google, Intel, and Amazon file the most amount of applications. Correspondingly, California where most technology companies are situated is on top of number of applications by states. We see a drop in number of applications after the 2008 crisis. It resumes high in 2010, but then steadily increases until 2013. The recent peak is in 2016, but the number again drops by almost twenty five percent in 2017 as the Trump administration issued tighter regulations on H-1B applications. Thus, the application is highly affected by economic cycle and policies. Different sectors share similar pattern in increase and drop in number of applications each year. 
\begin{figure}
  
    \begin{subfigure}[b]{0.5\textwidth}
        \includegraphics[width=\linewidth]{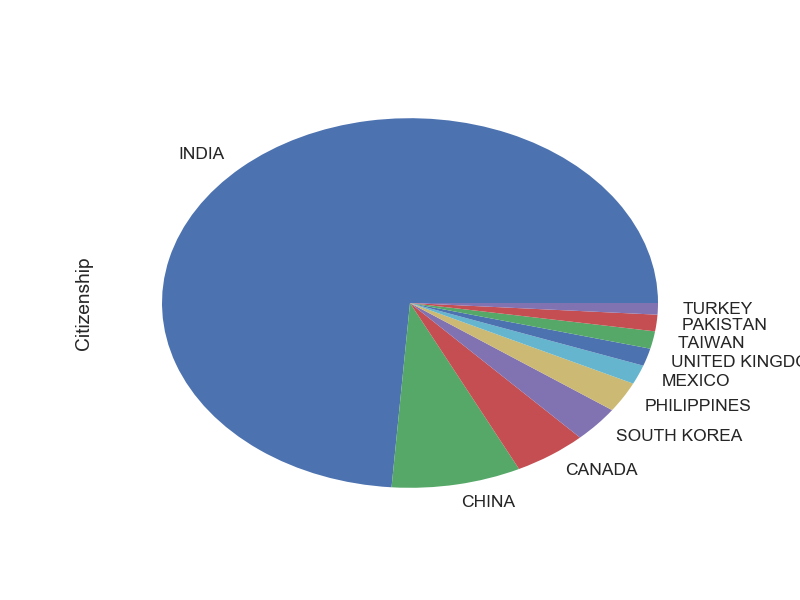}
        \caption{Number of applications by citizenship}
    \end{subfigure}
    \begin{subfigure}[b]{0.5\textwidth}
        \includegraphics[width=\textwidth]{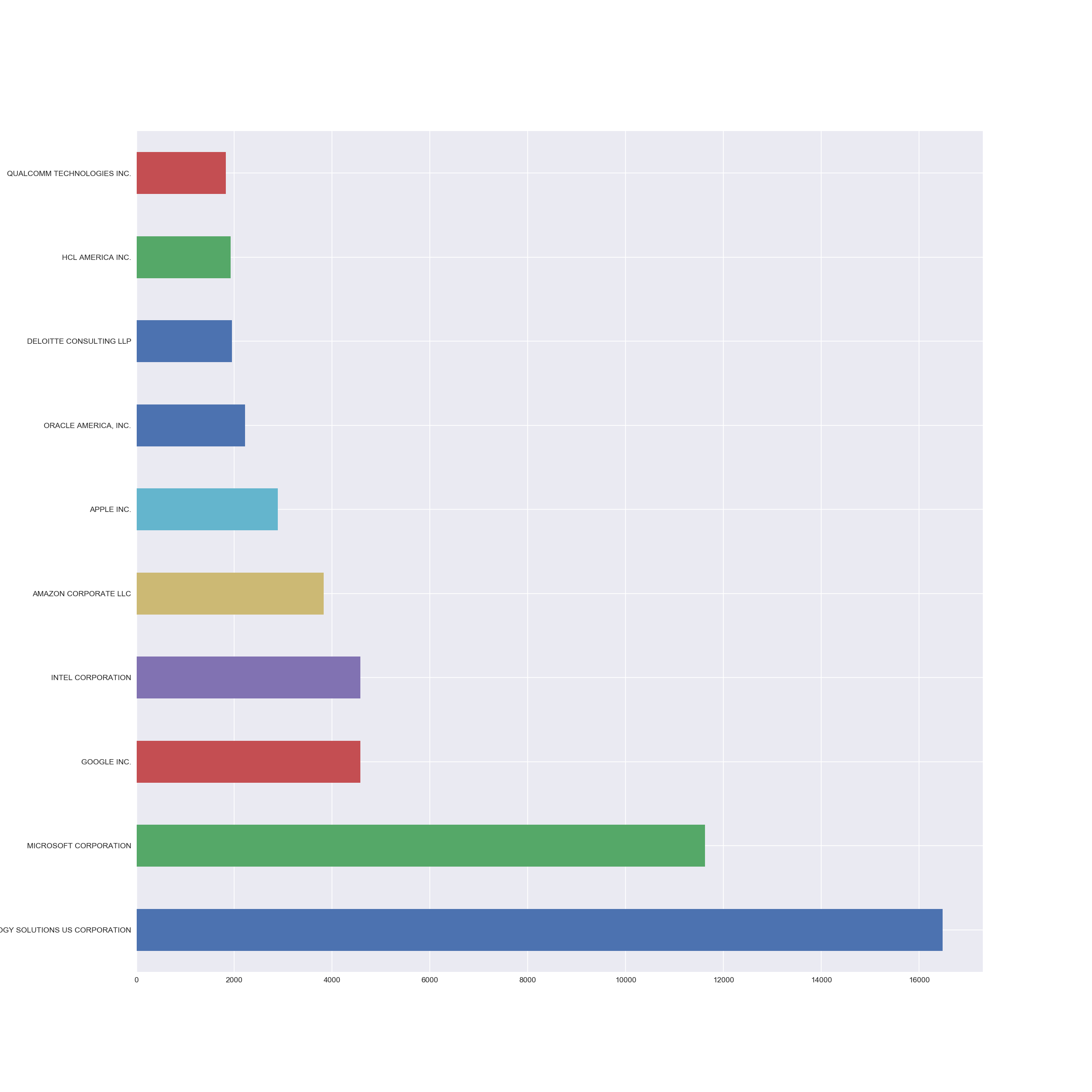}
        \caption{Number of applications by firm}
    \end{subfigure}
    
    \begin{subfigure}[b]{0.5\textwidth}
        \includegraphics[width=\textwidth]{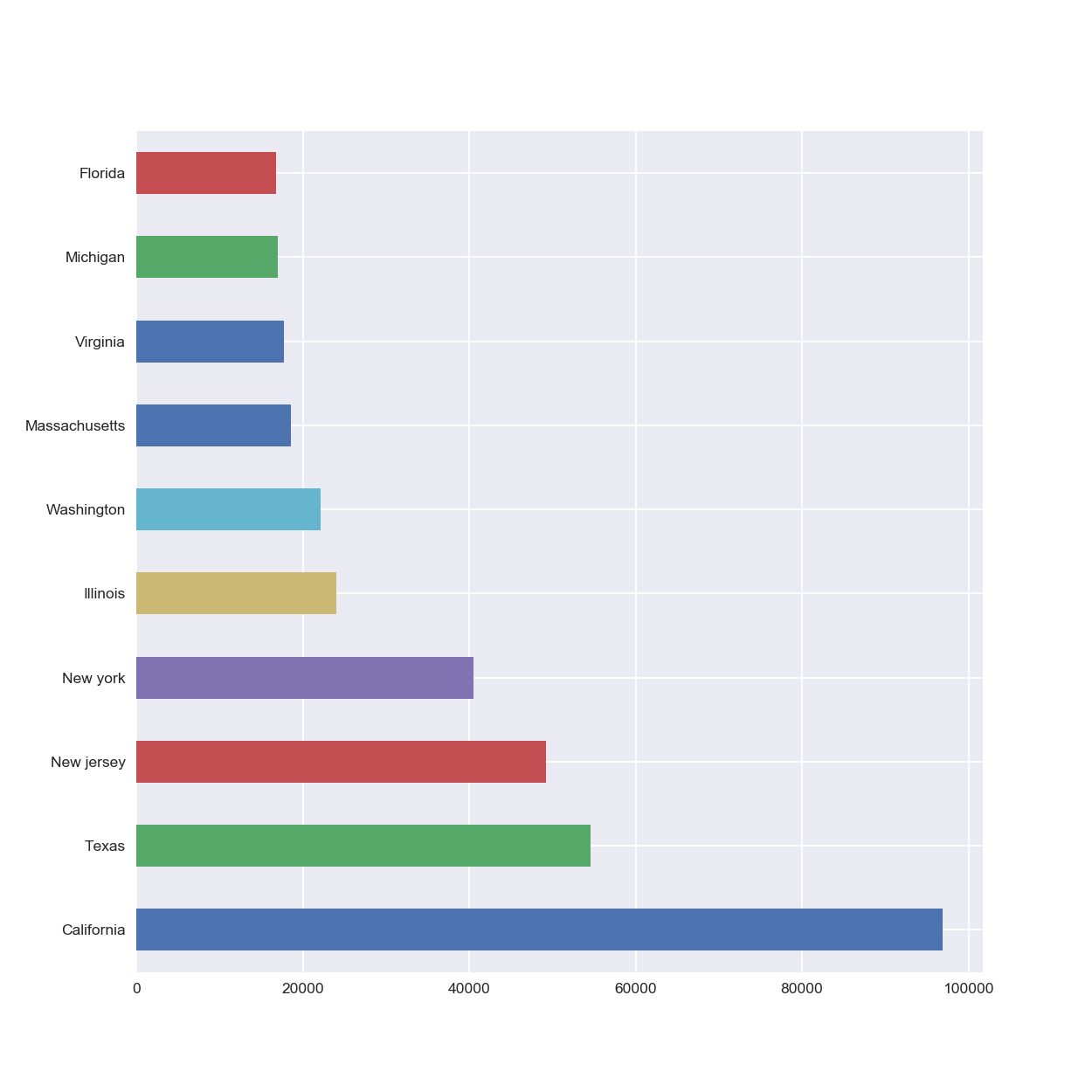}
        \caption{Number of applications by states}
    \end{subfigure}
    \begin{subfigure}[b]{0.5\textwidth}
        \includegraphics[width=\textwidth]{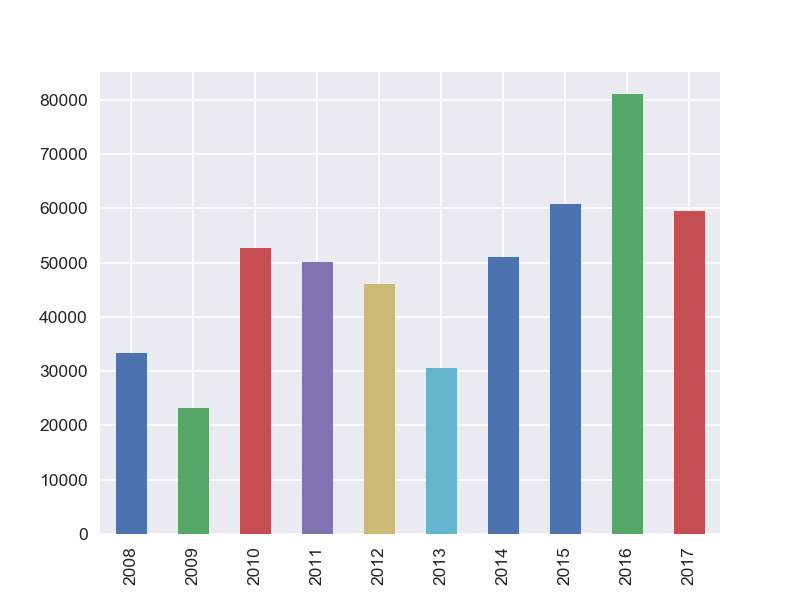}
        \caption{Number of applications by year}
    \end{subfigure}
    \caption{Box plot of wages}
\end{figure}

\section{Regression Analysis on Wage}

The first data analysis question is about the determinants of H-1B wages. The richness of cross-sectional features in the H-1B dataset provides us with great opportunities to understand how different employers set wages for employees with different backgrounds. We are interested in extracting the firm-level and individual-level features that have a significant impact on the wages given by the H-1B employers. Although we can achieve this by a simple Ordinary Least Square (OLS) regression and simply compare the coefficient for each feature, this may raise two potential issues.  

The first issue is prediction accuracy: the OLS estimates often have a low bias (or no bias) but large variance. Prediction accuracy can sometimes be improved by shrinking or setting some coefficients to zero. By doing so we sacrifice a little bit of bias to reduce the variance of the predicted values and hence may improve the overall prediction accuracy. The second issue is interpretability: with a large number of predictors, we often would like to determine a smaller subset that exhibits the strongest effects. Hence, in order to preserve both model performance and the interpretability of the model, for the wage regression, we propose to use Least Absolute Shrinkage and Selection Operator (LASSO) as our model to perform feature selection task for H-1B wages. 

\subsection{LASSO}

For each H-1B case $i$, we define $y_i$ to be the wage of this H-1B case, $x_i^j$ to be the value of feature $j$ of it and $x_i$ to be the vector of feature values of case $i$. Let $p$ be the total number of features and let $\beta = (\beta_1, \beta_2, \cdots \beta_p)$.
The model solves the following problem (\cite{main}, \cite{pathwise})
\begin{equation}\label{lasso}
        \min_{(\beta_0,\beta) \in \mathbb{R}^{p+1}} R_\alpha (\beta_0, \beta) = \min_{(\beta_0, \beta) \in \mathbb{R}^{p+1}} [\frac{1}{2N} \sum_{i=1}^N (y_i - \beta_0 -x_i^T\beta)^2 + \alpha P(\beta)]
\end{equation}

where $P(\beta)$ is the regularization term. LASSO Regression is a specific case where the regularization term represents the L1-norm:

\begin{equation}
        P(\beta) = ||\beta||_{\mathbb{L}_1} = \sum_{j=1}^p |\beta_j|
\end{equation}
To solve the optimization problem in LASSO we will use Coordinate Descent (CD) algorithm. It is not the paper's main objective to derive the functionality of CD algorithm, but the basic idea of Coordinate Descent is that we partially optimize with respect to one coordinate, assuming other coefficients are known at the optimum. Specifically, suppose we have estimates $\tilde{\beta_0}$ and $\tilde{\beta_l}$ for $l\neq j$ and we wish to partially optimize with respect to $\beta_j$. We want to take the gradient at $\beta_j=\tilde{\beta_j}$. Because of the L1 penalty term, it only exists if $\tilde{\beta_j} \neq 0$. The gradient of $R_\alpha(\beta_0,\beta)$ is

    \begin{equation}
        \frac{\partial R_\alpha}{\partial \beta_j}|_{\beta = \tilde{\beta}} = -\frac{1}{N} \sum_{i=1}^N x_{i}^j(y_i -\tilde{\beta_0} - x_i^T\tilde{\beta}) + \alpha    \end{equation}
if $\tilde{\beta_j} > 0$, and
    \begin{equation}
        \frac{\partial R_\alpha}{\partial \beta_j}|_{\beta = \tilde{\beta}} = -\frac{1}{N} \sum_{i=1}^N x_{i}^j(y_i -\tilde{\beta_0} - x_i^T\tilde{\beta}) - \alpha
    \end{equation}
if $\tilde{\beta_j} < 0$. By setting the gradient to 0, we can solve for the update scheme for $\tilde{\beta_j}$: 

\begin{equation}
    \tilde{\beta_j} \leftarrow S(\frac{1}{N}\sum_{i=1}^N x_i^j(y_i - \tilde{y_i}^{(j)}),\alpha)
\end{equation}
where $y_i - \tilde{y_i}^{(j)}$ is the partial residual of fitting $\beta_j$ and $S(z,\gamma)$ is the soft-thresholding operator with value $sign(z)(|z|-\gamma)_{+}$
The benefit of LASSO regression, as we can see from the update scheme, is that many features are set exactly at 0 for updating, and therefore it automatically performs feature selection while solving the optimization problem. 
In our project, we will primarily rely on the \texttt{scikit-learn} package in Python (\cite{IML}), which has built-in functions for solving LASSO problem using Coordinate Descent. 

\subsection{Fitting Real Data}
As mentioned in Section \ref{EDA}, the dataset contains more features for H-1B cases after 2015. Therefore, we will fit two models. Model 1 fits a LASSO regression using all the cases from 2008 to 2017 but only using a subset of all the features which are presented before 2015. These features include the economic sectors of the employer, the state of the employer, the citizenship of the applicants, the job level, the pay unit\footnote{Hourly, daily, weekly, monthly or annual payment. This is a reference for the length of the contract. A long-term job contract usually features a longer pay unit.}, and the year of the application. Model 2 fits a LASSO regression using the expanded feature space that is only presented in the dataset from 2015 to 2018. The features include all the features in Model 1 plus the major of the H-1B applicants in college, the education level, the ownership interest of the applicant\footnote{whether the applicant owns the company}, prior job experience as the number of months worked, the number of founding years, and the employer's total  number of workers.

As for data preprocessing, since many of the features are presented as categorical data, we will perform one-hot mapping for all the categorical features. And in order to limit the number of features after one-hot mapping, we will limit the one-hot mapping to the top 100 categories in each feature in order to reduce the feature space\footnote{This is especially relevant for LASSO regression since when $p>n$, LASSO selects at most $n$ features.}. Also note that we are including the application year as a feature for wage determination because we observed wage trends in time, so including year as a feature will serve as a "fixed time effect" for the model. 

For the LASSO regression, $\alpha > 0$ is the regularization parameter that controls the amount of shrinkage: the larger the value of $\alpha$, the greater the amount of shrinkage. We use 10-fold Cross Validation to select $\alpha$. Specifically, we divide the dataset for Model 1 and Model 2 into 10 parts. For the $k^{th}$ part $(k \in \{1,2,\cdots 10\})$, we fit the model to other 9 parts of the data and calculate the prediction error of the fitted model when predicting the $k^{th}$ part of the data. We do this for $k=1,2,\cdots 10$ and combine the $K$ estimates of prediction error. We denote the fitted function by $\hat{f}^{k(i)}(x_i, \alpha)$ which is the fitted value with the $k^{th}$ part of the data removed, and evaluate at $x_i \in R^p$. The cross-validation estimate of prediction error for this model is 

\begin{equation}
    CV(\hat{f},\alpha) = \frac{1}{n}\sum_{i=1}^n (y_i - \hat{f}^{k(i)}(x_i, \alpha))^2
\end{equation}
The function $CV(\hat{f},\alpha)$ provides an estimate of the test error given $\alpha$ and we hence find the tuning parameter $\hat{alpha}$ using grid-search from $\alpha \in \{\alpha_0, \alpha_1 \cdots \alpha_k\}$ that minimizes it. Our final chosen model is $f(x, \hat{\alpha})$ which we then fit to all data. 

We also perform the out-of-sample analysis for the trained model. Specifically, we randomly split the dataset into two parts: a training set which takes $\frac{4}{5}$ of the data, and a test set which takes the remaining $\frac{1}{5}$ of the data. We fit the LASSO Regression to the training data and tune the parameter $\alpha$ using the 10-fold Cross-Validation as described before. After we train the model, we apply it to the test set and attain a predicted value $\hat{y_i}$. We then calculate the out-of-sample $R^2$ (denoted as $R^2_{OS}$, as the following:

\begin{equation}
    R_{OS}^2 = \frac{[\sum_{i\in I_{OS}} ((\hat{y_i} - \bar{\hat{y_i}})(y_i - \bar{y_i})]^2}{\sum_{i\in I_{OS}} (\hat{y_i} - \bar{\hat{y_i}})^2 \sum_{i\in I_{OS}} (y_i - \bar{y_i})^2}
\end{equation}

\begin{figure}
    \begin{subfigure}[b]{0.5\textwidth}
        \includegraphics[width=\linewidth]{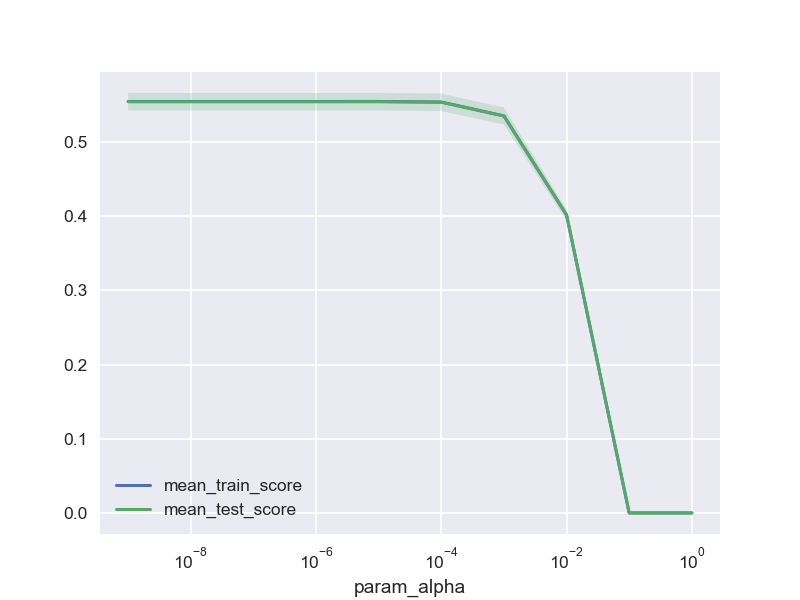}
        \caption{Model1: Data for features exist before 2015}
    \end{subfigure}
    \begin{subfigure}[b]{0.5\textwidth}
        \includegraphics[width=\linewidth]{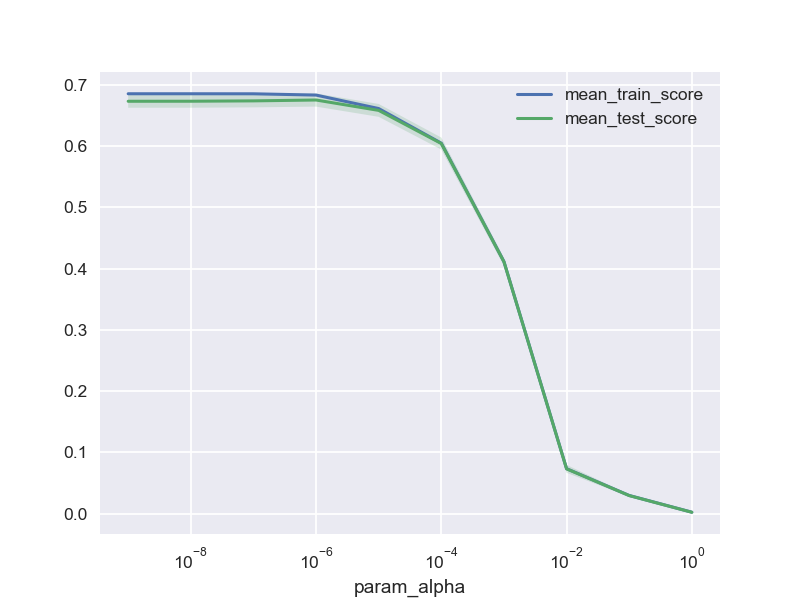}
        \caption{Model1: Data for features exist after 2015}
    \end{subfigure}
\caption{LASSO Cross-validation Result - CV Score $\pm$ Standard Error vs. Value of $\alpha$}\label{CV_lasso}
\end{figure}

The Cross-Validation result is shown in Figure \ref{CV_lasso}. Using grid-search we find the best tuning $\alpha = 1.0\times 10^{-6}$. We obtain an in-sample $R^2$ of 0.54 and out-of-sample $R^2$ of 0.54 for Model 1, and a in-sample $R^2$ of 0.67 and out-of-sample $R^2$ of 0.68 for Model 2. The similarity between in-sample and out-of-sample result shows no significant heterogeneity in cross-sectional prediction.

\subsection{Feature Importance}

\begin{figure}
    \centering
    \includegraphics[width = 0.8\textwidth]{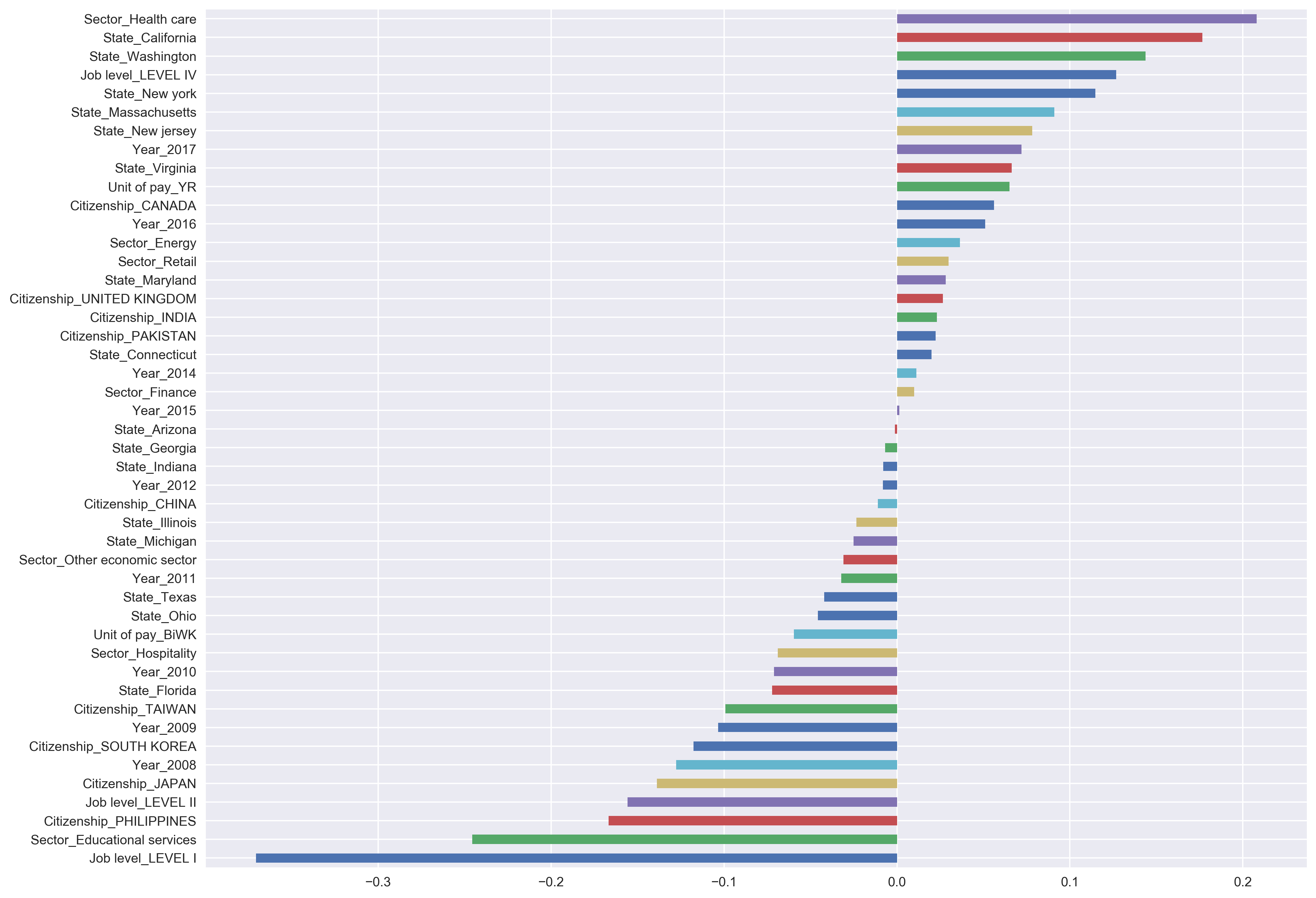}
    \caption{Feature Importance for Model 1: Only Pre-2015 Features}
    \label{fig:feature_wage_pre}
\end{figure}

\begin{figure}
    \centering
    \includegraphics[width = 0.8\textwidth]{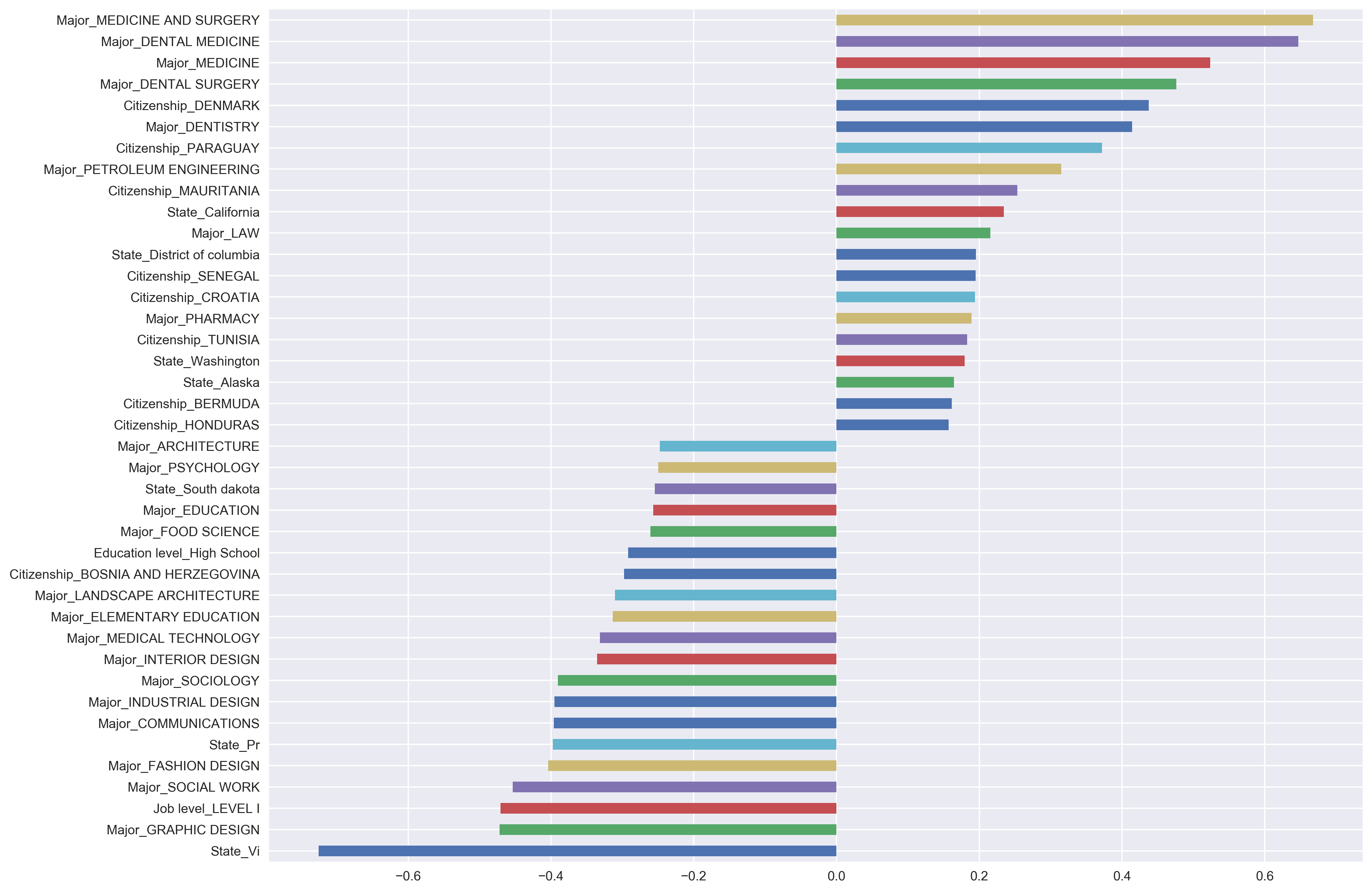}
    \caption{Feature Importance for Model 2: Both Pre-2015 and Post-2015 Features}
    \label{fig:feature_wage_post}
\end{figure}

The linear feature selection model helps us to select the most influential features that impact the wage of H-1B applicants. We look at both features that have the most positive coefficients and features with the most negative coefficients, so that we obtain a more comprehensive view of how different firm-level and individual-level features give both more and fewer wages. Figure \ref{fig:feature_wage_pre} and Figure \ref{fig:feature_wage_post} plot respectively the 20 most positive and 20 most negative features for Model 1 and Model 2. 

From the feature importance in Model 1 we conclude that H-1B applicants working in the healthcare industry usually receive higher wages. Also, working in California, Washington, New York, and Massachusetts usually yields higher wages. A higher job level (i.e. Job Level IV) also has a positive impact on the wage H-1B applicants get. On the other hand, lower job level (i.e. Job Level I and II) gives lower wages; working in education services gives lower wages, and having citizenship of Philippines or Japan also have a negative correlation with the wage. We also find that Year-2018 has a very negative coefficient, which is consistent with the finding of an increasing trend in wage, and the particular negative coefficient for 2008 is most likely due to the financial crisis.

Model 2 considers the additional features added after 2015 and therefore only uses data from 2015 to 2018. From the feature importance in Model 2 we find that H-1B applicants with majors in Medicine, surgery or pharmacy have higher wages, which is correlated to the high earnings found in the healthcare industry in Model 1. In addition, petroleum engineering and law major applicants also tend to have higher wages. As for the negative territory, we find majoring in graphic design, social work, fashion design, or communication usually have a negative impact on the wage.

\section{Classification Analysis of Case Status}

Another very relevant issue regarding H-1B application is whether the H-1B case gets approved by the U.S. Department of Labor. Each year the Department of Labor reviews the information of each application and decides if the case is certified or not. After being certified, the H-1B case will enter a lottery pool where an ex-ante random selection is performed to draw the cases to be finally approved. Our data from LCA covers only the first stage of deciding whether the case gets certified into the lottery pool, and we cannot observe if the H-1B cases are getting finally approved from the lottery or not. However, the random selection happening during the lottery should be non-discriminating and therefore should not change the ex-post distribution of the approved H-1B cases. Therefore, we assume that the feature selections, as well as prediction of the certification during the first stage, will also speak to the final approval of the H-1B case. 

We formulate this analysis as a classification problem. For each H-1B case, we observe an outcome, whether being positive (\texttt{Certified or Certified-Expired}) or negative (\texttt{Denied}). We will train a classification model to predict whether a given H-1B case will be certified or not, and what are the features that contributed the most to the certification/denial decision of the H-1B case. Similar to our analysis for H-1B wages, we are concerned with both model prediction and model interpretability. For better model prediction we need to account for the bias-variance tradeoff in model selection, and for better interpretability, we need a parametric model whose estimated parameters can make economic sense. As a result of such consideration, we propose to use Logistic Regression with L1-Penalty, a cousin of the LASSO regression used in our analysis in Section 3, as our model for the classification analysis.

\subsection{Logistic Regression with L1-Penalty}

For each case $i$, we define its case status be $y_i = 1$ if it is certified or $y_i = 0$ if it is not. Suppose $x_i^j$ is the value of $j^{th}$ feature of case $i$ and $x_i$ be the feature vector. The unpenalized logistic regression model takes the following functional form

\begin{align}
    Pr(y_i = 1|x_i) &= \frac{1}{1+e^{-(\beta_0 + x_i^T \beta})}     \\
    Pr(y_i = 0|x_i) &= 1-Pr(y_i = 1|x_i) = \frac{1}{1+e^{(\beta_0 + x_i^T \beta)}}
\end{align}

Let $p(x_i) = Pr(y_i = 1|x_i)$, we want to maximize the log likelihood of the joint distribution. Similar to LASSO regression, we add an L1 penalty term to the log-likelihood function for regularization. The maximization problem is (\cite{main})

\begin{equation}
    \max_{(\beta_0,\beta)\in \mathbb{R}^{p+1}} [\frac{1}{N}\sum_{i=1}^N \{y_i \log p(x_i) + (1-y_i)\log(1-p(x_i))\} - \lambda P_{\mathbb{L}_1}(\beta)]
\end{equation}

where $P_{\mathbb{L}_1}(\beta)$ is the L1-norm. Again, there are many ways to solve this optimization problem and the way how it is implemented is not the main focus of our paper. However, it's worth mentioning the basic numerical treatment just for reference, because it is very hard to perform the Coordinate Descent algorithm for this problem since the gradient solution yields no analytic solution. Instead, we focus on an approximated problem. Consider the unpenalized log-likelihood function $\ell(\beta_0, \beta)$ as 

\begin{equation}
    \ell(\beta_0, \beta) = \frac{1}{N} \sum_{i=1}^N y_i (\beta_0 + x_i^T \beta) - log(1+e^{(\beta_0 + x_i^T\beta)}
\end{equation}

Denote $\tilde{\beta_0}$ and $\tilde{\beta}$ be the current estimates, if we perform Taylor expansion of the unpenalized log-likelihood function around $\tilde{\beta_0}$ and $\tilde{\beta}$ to the second order we will get a quadratic approximation 

\begin{equation}
    \ell_Q(\beta_0,\beta) := -\frac{1}{2N} \sum_{i=1}^N w_i (z_i - \beta_0 -x_i^T\beta)^2 + {\cal O}(||\beta-{\tilde{\beta}}||^2)
\end{equation}

where

\begin{align}
    z_i &= \tilde{\beta}_0 + x_i^T \tilde{\beta} + \frac{y_i - \tilde{p}(x_i)}{\tilde{p}(x_i)(1-\tilde{p}(x_i))} \\
    w_i &= \tilde{p}(x_i)(1-\tilde{p}(x_i))
\end{align}

Then we can write the approximated optimization problem for Logistic Regression with L1-Penalty to be 

\begin{equation}
    \min_{(\beta_0,\beta)\in \mathbb{R}^{p+1}} [-\ell_Q(\beta_0,\beta) + \lambda P_{\mathbb{L}}(\beta)]
\end{equation}
We can see that such minimization problem is the same as the one in LASSO regression, and therefore by performing the Coordinate Descent update scheme we can easily solve the problem. As another benefit from this, we also obtain many feature weights to be set exactly at zero, which performs feature selection. In our analysis this part, we also rely on the built-in functions in \texttt{scikit-learn} package to perform model training.  

\subsection{Fitting Real Data}
In order to utilize the richness of cross-sectional features, we will perform the classification analysis on the dataset from 2015 to 2018, where more features are added. Similar to Model 2 in the Section 3, the features we consider for our classification model includes wage, economic sector, job level, pay unit, working state (location), education level, job experience, applicants college major, employer's history, ownership interest, and the total number of employees of the firm. Similarly, we use one-hot mapping for the 100 most frequent categories for the categorical features. As for model evaluation, we use the Area Under the Receiver Operating Characteristic Curve (AUC) as our metrics following standard literature for classification problem \cite{roc}. The Receiver Operating Characteristic (ROC) is created by plotting the true positive rate (TPR) against the false positive rate (FPR) at various threshold settings, and a larger area under the ROC curve means more distinguished classes and a better classifier. 

When training the classifier, we want to make sure that both positive classes and negative classes have enough representations in the sample. That is not the case for original sample, where among the total 191693 H-1B cases from 2015 to 2018, only 7385 - about 5\%  cases are denied. Since the dataset is very unbalanced, we perform 1) undersampling from positive class and 2) oversampling from negative class to generate the training set so that the number of two classes are the same. 

For the Logistic Regression with L1-Penalty, $\lambda > 0$ is the regularization parameter that controls the amount of shrinkage. We use 10-fold Cross-Validation to select $\lambda$ similar to what we did in LASSO regression in Section 3. We also perform an out-of-sample test, a test on 20\% randomly selected sample from original dataset, for prediction performance using AUC as the measurement. The entire analysis scheme can be summarized as the following:

\singlespacing
\begin{itemize}
    \item Randomly select test set and training set
    \item Generate a list of $\{\lambda_1,\lambda_2, \cdots, \lambda_n\}$
    \item For each $\lambda_i$:
    \begin{itemize}
        \item Generate oversampling from denied class in training set, call it Sample 1
        \item Generate undersampling from certified class in training set, call it Sample 2
        \item Use the original training set as Sample 3
        \item For each sample in Sample 1 to 3:
        \begin{itemize}
            \item Use coordinate descent method to compute MLE of penalized LogReg with quadratic approximation of each sample
            \item Calculate AUC for each model
        \end{itemize}
        \item Calculate average AUC and call it the score of $\lambda_i$
    \end{itemize}
    \item Find the $\lambda_i$ with the highest average AUC
\end{itemize}
\onehalfspacing

\begin{figure}
    \centering
    \includegraphics[width = \textwidth]{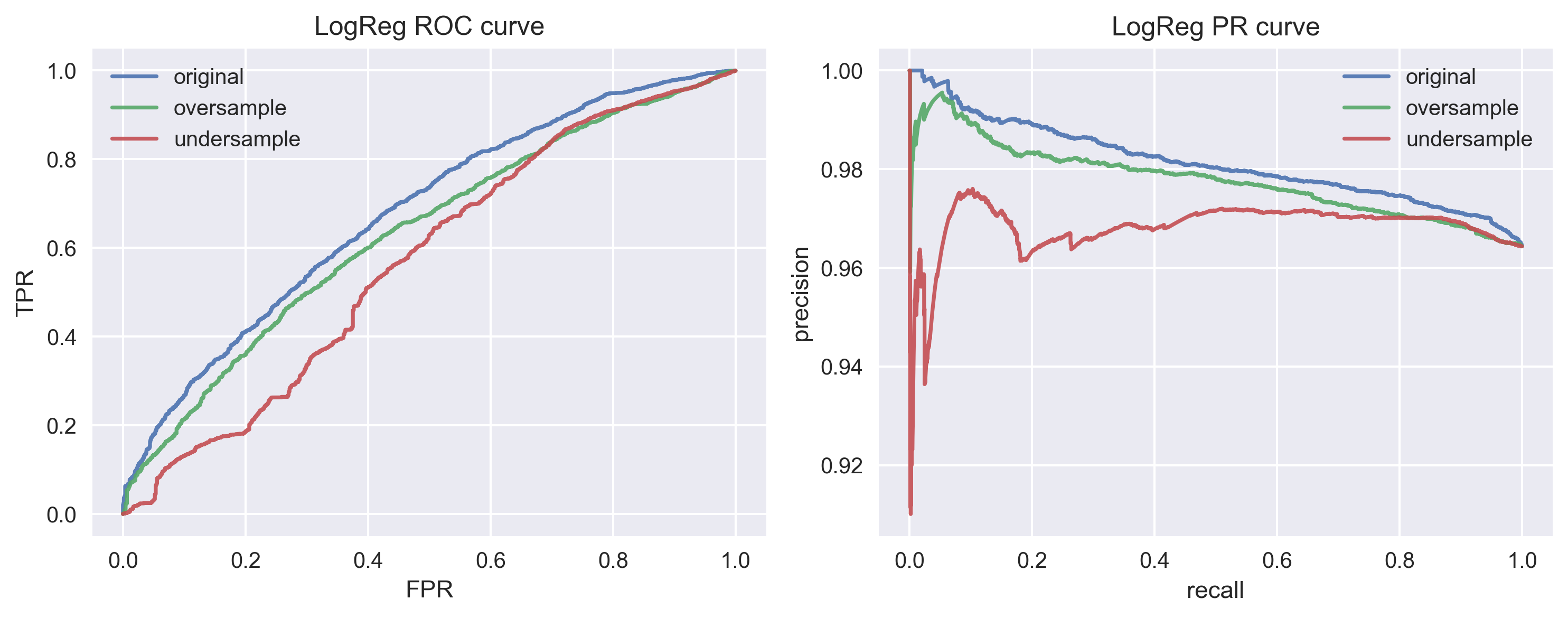}
    \caption{Out-of-sample ROC curve and Precision-Recall curve for the trained model with different random sampling scheme}
    \label{fig:roc}
\end{figure}

The grid-search gives optimal $\lambda = 1$. Under this choice of $\lambda$, We obtained the highest in-sample AUC score of 0.674 and out-of-sample AUC score of 0.676 for Sample 3 (original sampling). Figure \ref{fig:roc} shows the out-of-sample performance of the trained model under the best choice of $\lambda$.

\subsection{Feature Importance}

\begin{figure}
    \centering
    \includegraphics[width = \textwidth]{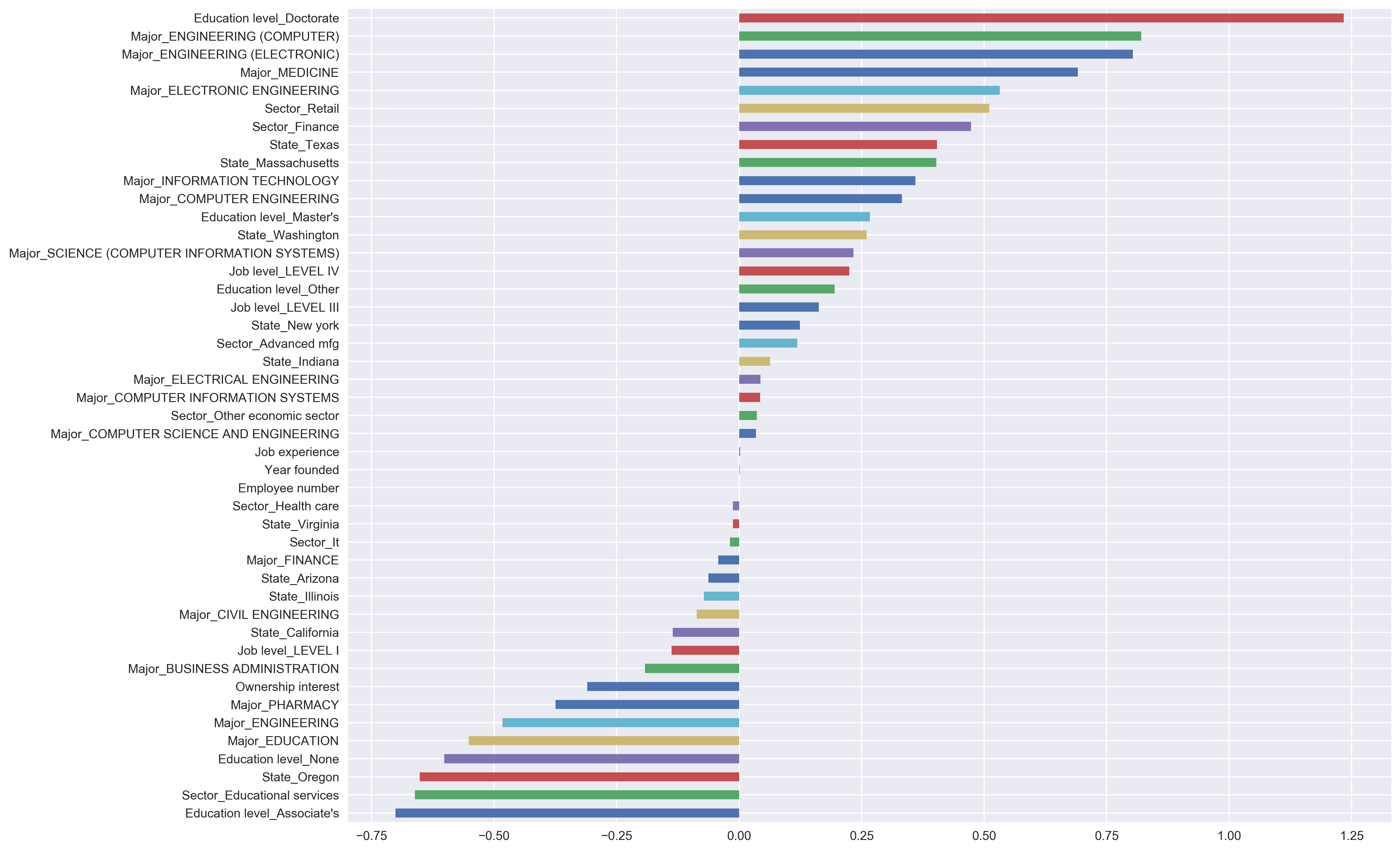}
    \caption{Feature Importance from Logistic Regression with L1-Penalty model, with $\lambda = 1$ at original training dataset}
    \label{fig:featurelog}
\end{figure}

We look at both the positive and negative features selected from the Logistic Regression with L1-Penalty. Figure \ref{fig:featurelog} plots the coefficients for the most positive and negative features. We find that applicants having a Ph.D. degree, majoring in Computer Science or Electrical Engineering or Medicine, and working in Retail or finance sector are more likely to get their H-1B case certified, whereas applicants having no or only an Associate degree, majoring in Education, and working in the education services sector will have a higher chance to have their cases rejected.

\section{Discussion}

\subsection{Model Performance with Different Sampling Frequency}
One interesting point we notice during our classification analysis for H-1B certification is that despite we have a very unbalanced dataset, it still gives the best performance compared with the dataset generated from random sampling that has an equal number of representation from positive and negative class. This is contradicting our expectation that a classifier is trained to the best when the represented classes are of similar weights. To study the performance of models with different sampling frequency and to check the robustness of our trained model, we perform an additional test of the model with different sampling frequencies and different hyperparameter $\lambda$. The procedure follows 

\singlespacing
\begin{itemize}
    \item Generate a list of hyperparameter $\Lambda = \{\lambda_1, \lambda_2, \cdots \lambda_k\}$
    \item For each $\lambda \in \Lambda$:
    \begin{itemize}
        \item Generate a list of sampling frequencies $\gamma$ that maps to $[0,1]$, where $0$ means no random sampling (original sample) and $1$ means $50/50$ sampling (equal weights from two classes)
        \item For each $\gamma$
        \begin{itemize}
            \item  Construct oversampling sample from Denied class and undersampling sample from Certified class
            \item Together with the original sample, train the LogReg model with L1-Penalty with $\lambda$ being the regularization parameter
            \item Calculate out-of-sample AUC score
        \end{itemize}
        \item Plot the AUC score against sampling frequency $\lambda$
    \end{itemize}
\end{itemize}
\onehalfspacing

\begin{figure}
  \centering
    \begin{subfigure}[b]{0.45\textwidth}
        \includegraphics[width=\linewidth]{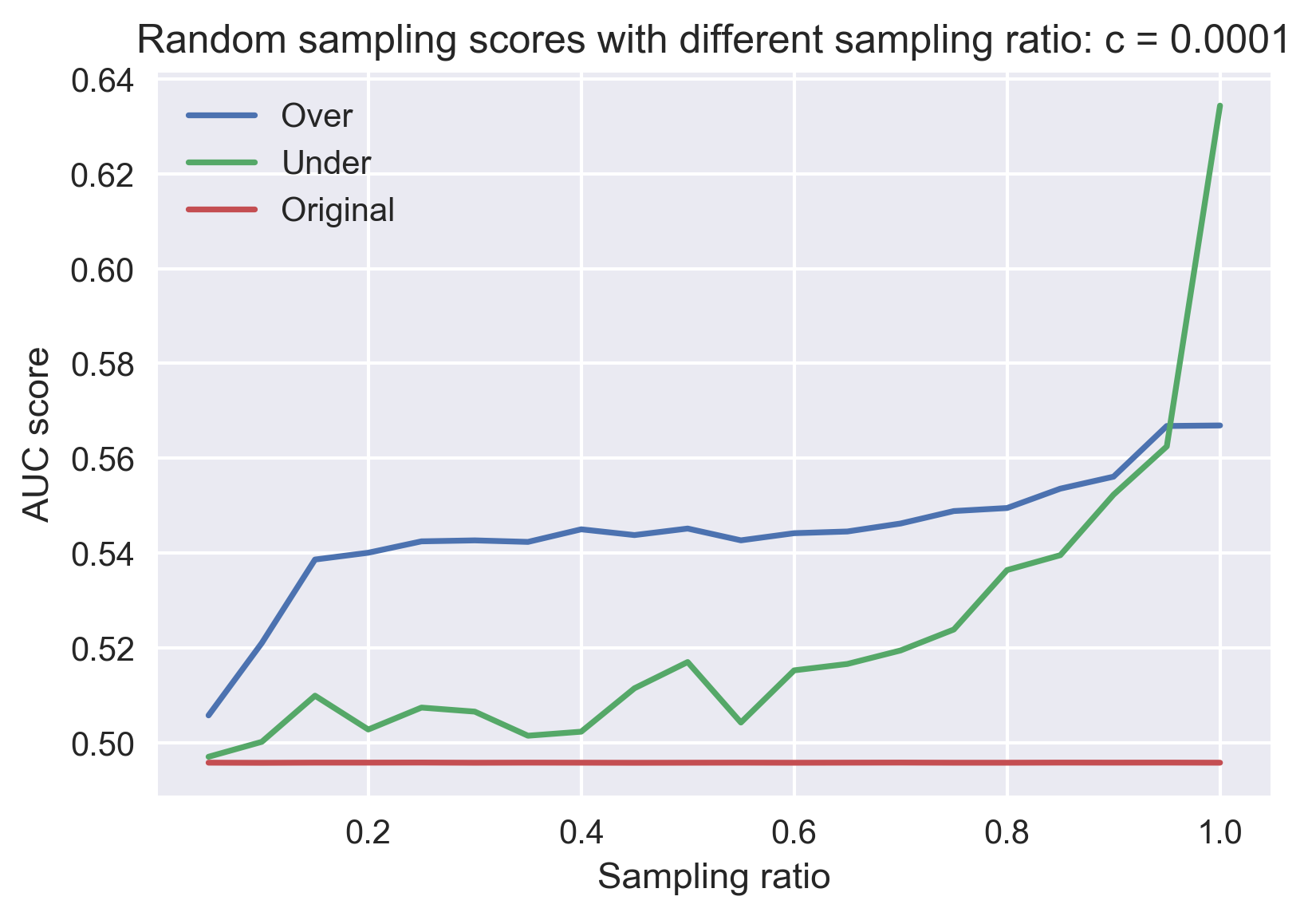}
    \end{subfigure}
    \begin{subfigure}[b]{0.45\textwidth}
        \includegraphics[width=\linewidth]{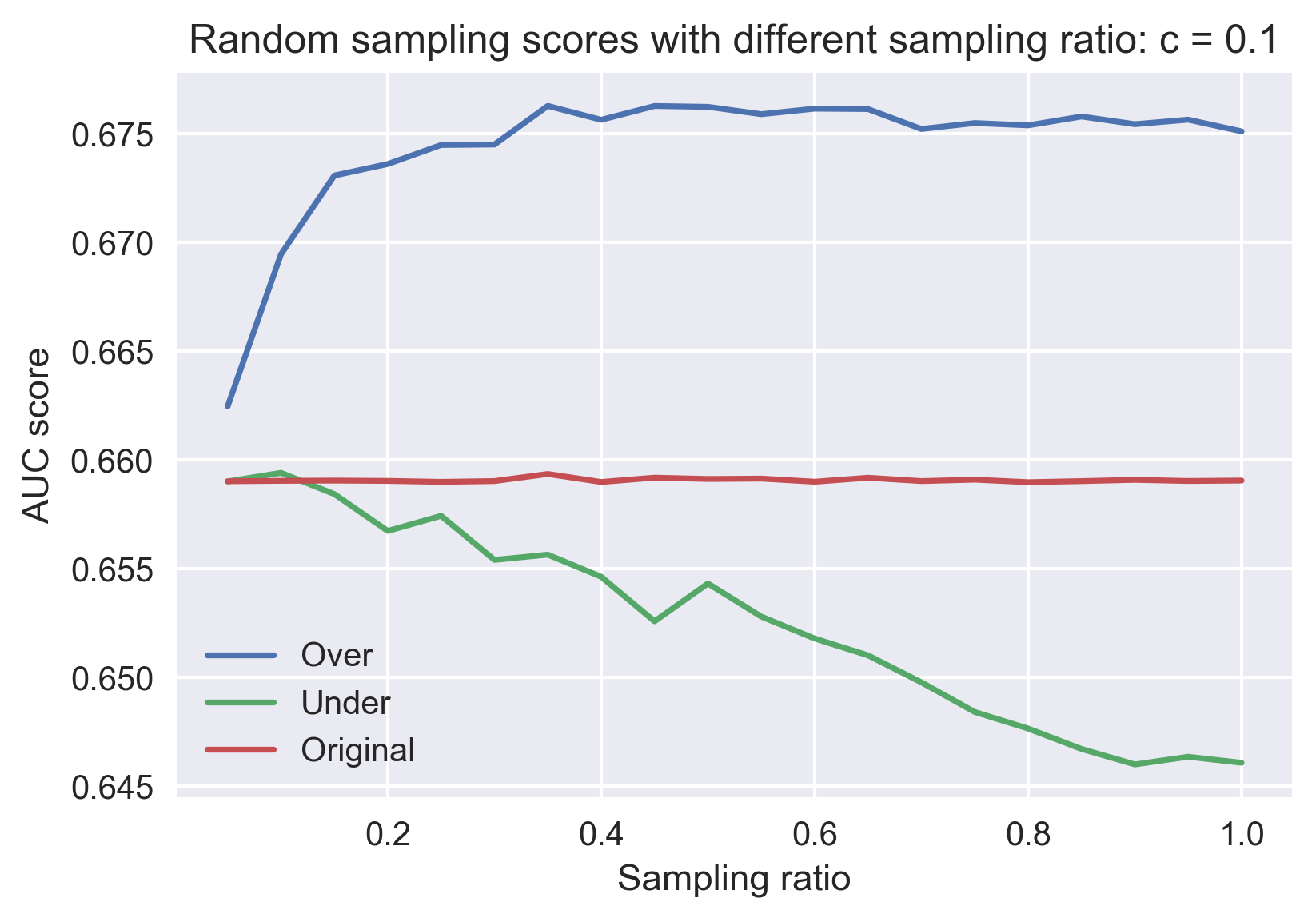}
    \end{subfigure}
    \\
    ~ 
    \begin{subfigure}[b]{0.45\textwidth}
        \includegraphics[width=\linewidth]{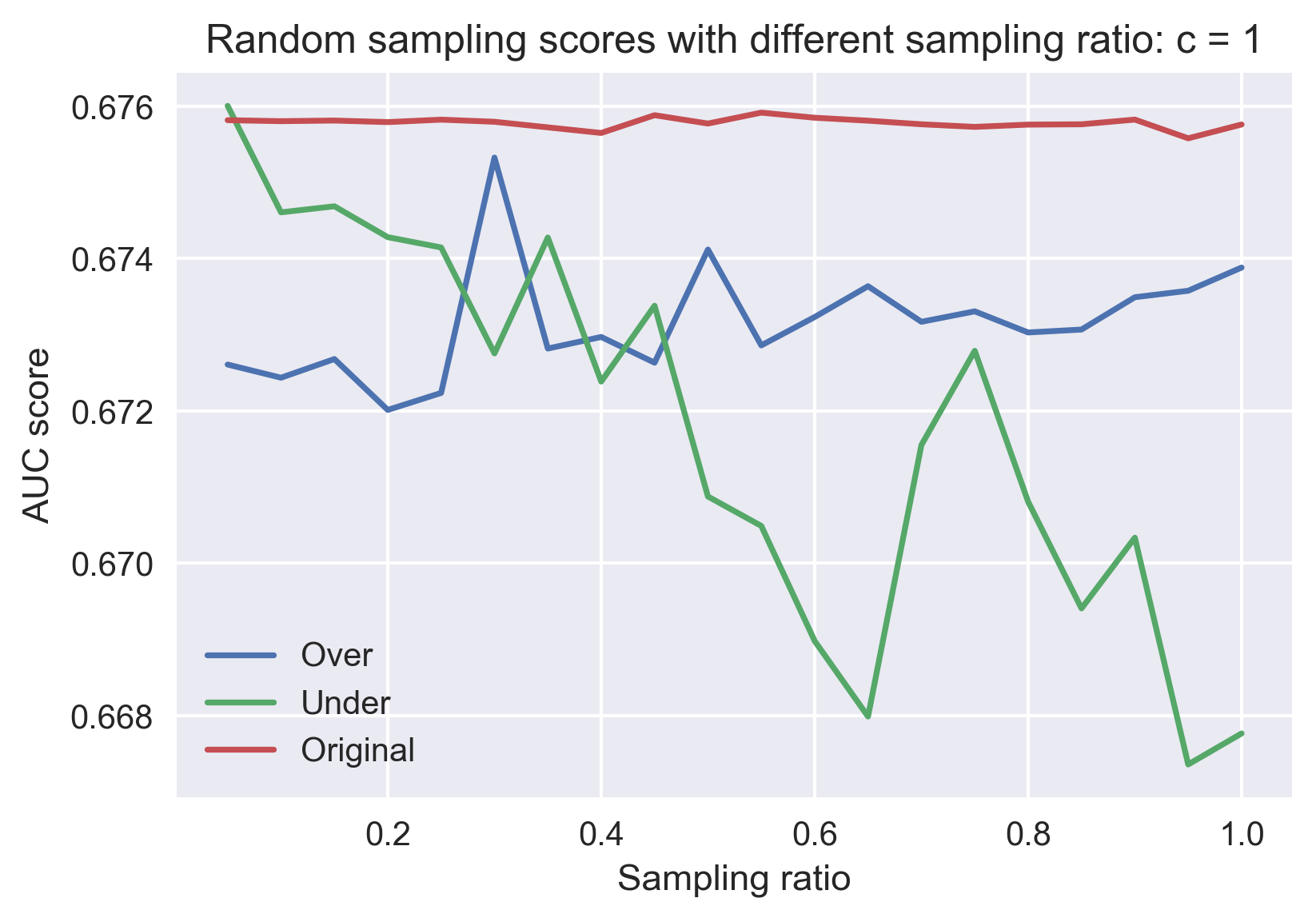}
    \end{subfigure}
    \begin{subfigure}[b]{0.45\textwidth}
        \includegraphics[width=\linewidth]{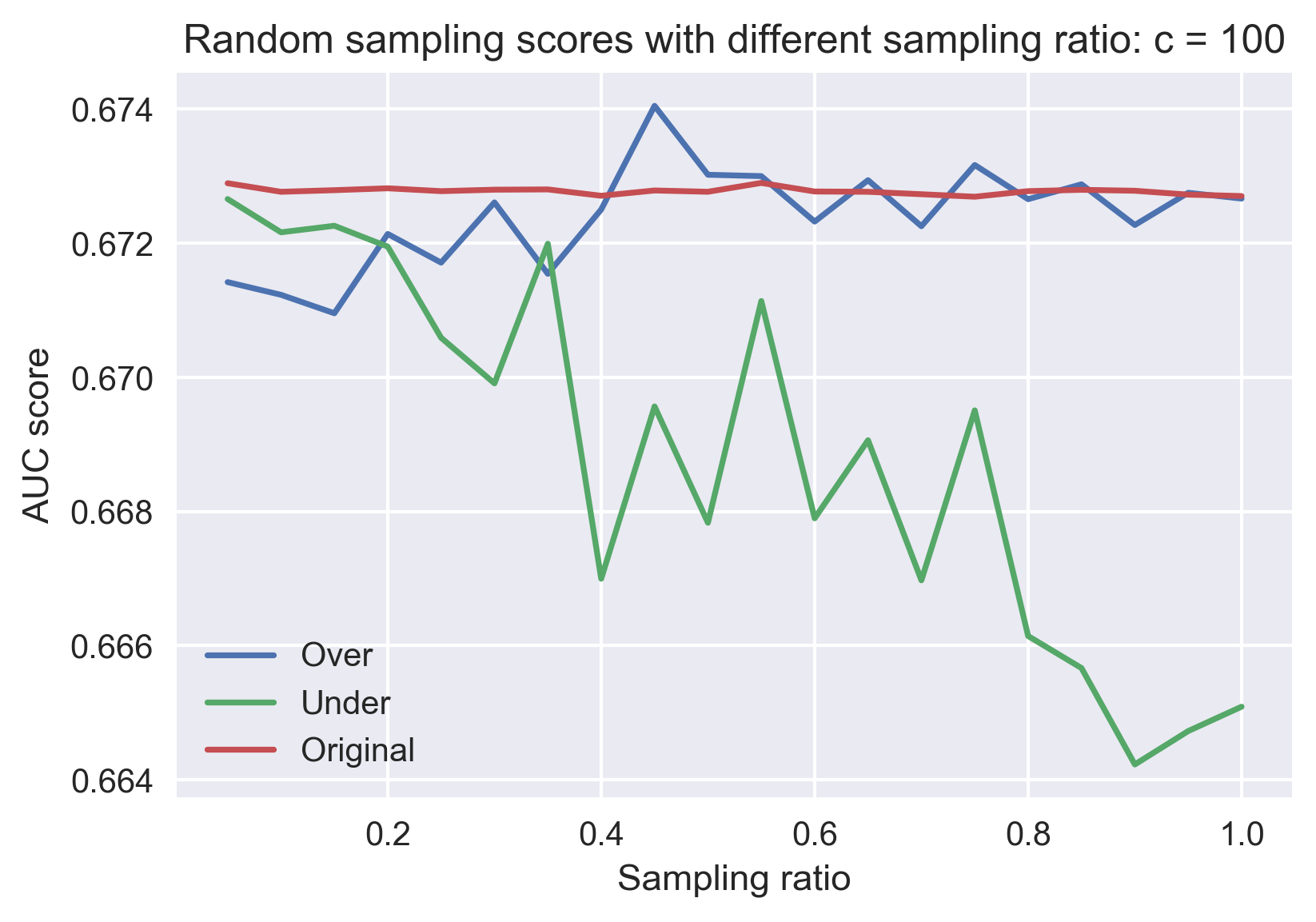}
    \end{subfigure}    
    \caption{Model performance under different sampling frequencies. 0 being no random sampling and 1 being 50/50 sampling frequency}\label{fig:robust}
\end{figure}

For simplicity we choose the set of hyperparameter $\Lambda = \{10^{-4}, 0.1, 1, 100\}$. Figure \ref{fig:robust} shows the plot of AUC under different sampling frequencies (0 being no random sampling and 1 being perfect sampling). We see that the performance of randomly sampled models changes dramatically for a different choice of hyperparameters, which challenges our notion that we should do random sampling as a necessary step for data pre-processing. On the other hand, the highest score obtained from the original sample in the $\lambda = 1$ plot checks robustness for our analysis in Section 4.

\subsection{Model Strength}
Now we would like to discuss a little bit of the strength and weakness of our analysis for both H-1B wage and status. For both analyses, the biggest strength of our model is that it preserves the best interpretability when modifying model performance. Both LASSO regression and Logistic Regression with L1-Penalty gives feature coefficients that can be directly compared with each other and be interpreted with economic intuition. In both cases, there is only one hyperparameter to tune and therefore avoids the threat of over-fitting. 

\subsection{Potential Improvements}
Nevertheless, our models and analysis can also be improved from different perspectives. The first and most obvious shortcoming is the lack of identification strategy that identifies a causal relationship. Our models do not rule out confounding factors that are not presented in the feature space but could potentially influence the outcome. Therefore the results presented in this paper can only be viewed as an application of machine learning that finds correlations in the real world, rather than rigorous economic research that pins down causality of how wage is determined by H-1B employers or how certification decision is made by the Department of Labor. A further improvement toward this direction is to implement less machine learning, but more econometric tools such as Instrumental Variables (IV) or, since we have a panel dataset, use Difference-in-difference (Diff-in-diff) method to identify causality.

Another potential improvement to our models is to find determinants of wage and H-1B status beyond cross-sectional variations. While it is very important to understand cross-sectional differences, people might also be interested in learning predictions in time-series, for example, how would the trend of wage increase continue, or whether the Department of Labor will look more at some characteristics in the future. A major obstacle for conducting such analysis is that H-1B applications are very subjective to exogenous policy shocks, and the distributions from year to year could be completely different due to some immigration policies implemented by the government. In our analysis, we controlled time-fixed effect, but a further step is, hopefully with more data, to look at how the implementation of different immigration policies can influence the H-1B applications.

\section{Conclusion}
In this paper, we apply machine learning approaches to study the H-1B application dataset. Use LASSO regression model we studied the features that have the most significant impact on the H-1B applicants' wages. We find that applicants working in the healthcare industry or majored in healthcare-related majors usually have the highest wages (detailed result in Section 3.3). We trained a Logistic Regression with L1-Penalty as a classifier to extract features that have the most impact on the likelihood of having H-1B application certified. We find applicants with higher education level and majoring in computer science/electrical engineering have higher possibility to have their application certified. We also find that the performance of the classifier is higher when not doing random sampling for the unbalanced dataset, which casts doubt on the widely held notion of using random sampling as a step of pre-processing and calls for case-by-case analysis for different datasets. 

\clearpage
\bibliographystyle{unsrt}
\bibliography{template}

\end{document}